\documentclass[aip,apl,reprint]{revtex4-1}

\usepackage{amsmath}               
\usepackage{siunitx}
\usepackage{graphicx}
\usepackage{hyperref}
\usepackage{color}

\usepackage{xr}
\begin{document}



\title{Compact 3D quantum memory} 



\author{Edwar Xie}
\email{edwar.xie@wmi.badw.de}
\affiliation{Walther-Mei\ss ner-Institut, Bayerische Akademie der Wissenschaften, 85748 Garching, Germany}
\affiliation{Physik-Department, Technische Universit{\"a}t M{\"u}nchen, 85748 Garching, Germany}
\affiliation{Nanosystems Initiative Munich (NIM), Schellingstra\ss e 4, 80799 M{\"u}unchen, Germany}
\author{Frank Deppe}
\affiliation{Walther-Mei\ss ner-Institut, Bayerische Akademie der Wissenschaften, 85748 Garching, Germany}
\affiliation{Physik-Department, Technische Universit{\"a}t M{\"u}nchen, 85748 Garching, Germany}
\affiliation{Nanosystems Initiative Munich (NIM), Schellingstra\ss e 4, 80799 M{\"u}unchen, Germany}
\author{Michael Renger}
\affiliation{Walther-Mei\ss ner-Institut, Bayerische Akademie der Wissenschaften, 85748 Garching, Germany}
\affiliation{Physik-Department, Technische Universit{\"a}t M{\"u}nchen, 85748 Garching, Germany}
\author{Daniel Repp}
\affiliation{Physik-Department, Technische Universit{\"a}t M{\"u}nchen, 85748 Garching, Germany}
\author{Peter Eder}
\affiliation{Walther-Mei\ss ner-Institut, Bayerische Akademie der Wissenschaften, 85748 Garching, Germany}
\affiliation{Physik-Department, Technische Universit{\"a}t M{\"u}nchen, 85748 Garching, Germany}
\affiliation{Nanosystems Initiative Munich (NIM), Schellingstra\ss e 4, 80799 M{\"u}unchen, Germany}
\author{Michael Fischer}
\affiliation{Walther-Mei\ss ner-Institut, Bayerische Akademie der Wissenschaften, 85748 Garching, Germany}
\affiliation{Physik-Department, Technische Universit{\"a}t M{\"u}nchen, 85748 Garching, Germany}
\affiliation{Nanosystems Initiative Munich (NIM), Schellingstra\ss e 4, 80799 M{\"u}unchen, Germany}
\author{Jan Goetz}
\affiliation{Walther-Mei\ss ner-Institut, Bayerische Akademie der Wissenschaften, 85748 Garching, Germany}
\affiliation{Physik-Department, Technische Universit{\"a}t M{\"u}nchen, 85748 Garching, Germany}
\author{Stefan Pogorzalek}
\affiliation{Walther-Mei\ss ner-Institut, Bayerische Akademie der Wissenschaften, 85748 Garching, Germany}
\affiliation{Physik-Department, Technische Universit{\"a}t M{\"u}nchen, 85748 Garching, Germany}
\author{Kirill G. Fedorov}
\affiliation{Walther-Mei\ss ner-Institut, Bayerische Akademie der Wissenschaften, 85748 Garching, Germany}
\affiliation{Physik-Department, Technische Universit{\"a}t M{\"u}nchen, 85748 Garching, Germany}
\author{Achim Marx}
\affiliation{Walther-Mei\ss ner-Institut, Bayerische Akademie der Wissenschaften, 85748 Garching, Germany}
\author{Rudolf Gross}
\email{rudolf.gross@wmi.badw.de}
\affiliation{Walther-Mei\ss ner-Institut, Bayerische Akademie der Wissenschaften, 85748 Garching, Germany}
\affiliation{Physik-Department, Technische Universit{\"a}t M{\"u}nchen, 85748 Garching, Germany}
\affiliation{Nanosystems Initiative Munich (NIM), Schellingstra\ss e 4, 80799 M{\"u}unchen, Germany}


\date{\today}

\begin{abstract}
Superconducting 3D microwave cavities offer state-of-the-art coherence times and a well controlled environment for superconducting qubits. In order to realize at the same time fast readout and long-lived quantum information storage, one can couple the qubit both to a low-quality readout and a high-quality storage cavity. However, such systems are bulky compared to their less coherent 2D counterparts. A more compact and scalable approach is achieved by making use of the multimode structure of a 3D cavity. In our work, we investigate such a device where a transmon qubit is capacitively coupled to two modes of a single 3D cavity. The external coupling is engineered so that the memory mode has an about 100 times larger quality factor than the readout mode. Using an all-microwave second-order protocol, we realize a lifetime enhancement of the stored state over the qubit lifetime by a factor of $6$ with a fidelity of approximately \SI{80}{\percent} determined via quantum process tomography. We also find that this enhancement is not limited by fundamental constraints. 

\end{abstract}


\maketitle 


Engineered quantum systems are a versatile resource for promising applications, such as quantum computing\cite{Bennett2000,Vandersypen2001}, quantum communication \cite{Furusawa1998, DiCandia2015} and quantum cryptography \cite{Ekert1991}. Commonly, quantum resources are fairly fragile and decay on a short timescale. Therefore, they have to be protected by placing them in a well-isolated enviroment \cite{Raimond2001}.
In circuit quantum electrodynamics (QED), this is typically realized by coupling a superconducting qubit to a microwave resonator serving both for qubit protection \cite{Paik2011} and readout \cite{Blais2004}. Moreover, the resonator itself can be designed to decay much slower than the qubit \cite{Wallraff2004}. Then, the lifetime of a quantum state can be enhanced by the intermediate storage of the qubit state in a long-lived resonator mode. In terms of coherence times, 3D microwave waveguide cavities yield the best results so far, reaching the millisecond regime \cite{Reagor2016,Rigetti2012}. Other quantum memories can be constructed using microscopic defect states \cite{Neeley2008,Kubo2011,Zhu2011} or a nano-mechanical oscillator \cite{Reed2017}. In circuit QED architectures, long storage times and fast readout are conflicting requirements, asking for slow and fast cavity decay, respectively. A possible solution is to physically separate the storage and the readout resonators \cite{Sirois2015,Vlastakis2013,Liu2017,Wang2017,Ofek2016}. However, regarding scalabitiy and reduction of footprint, this approach is not optimal. By combining the high $Q$-factors of 3D cavities with the fast readout capabilities and compactness of 2D resonators, the system dimensions can be reduced \cite{Axline2016}. Another promising approach is to use the multimode structure of 2D resonators \cite{Leek2010}. In this case, the system turns out to be even more compact and scalable, however, this is achieved at the expense of a rather low storage time. To combine the benefits from both approaches, the multimode structure of a single 3D cavity can be employed\cite{Gasparinetti2016,Flurin2017}. On the one hand, we take advantage of high-$Q$ 3D cavities and, on the other hand, we entirely get rid of a separate readout resonator without losing the fast readout capability. 

\begin{figure}
\includegraphics[width=\columnwidth]{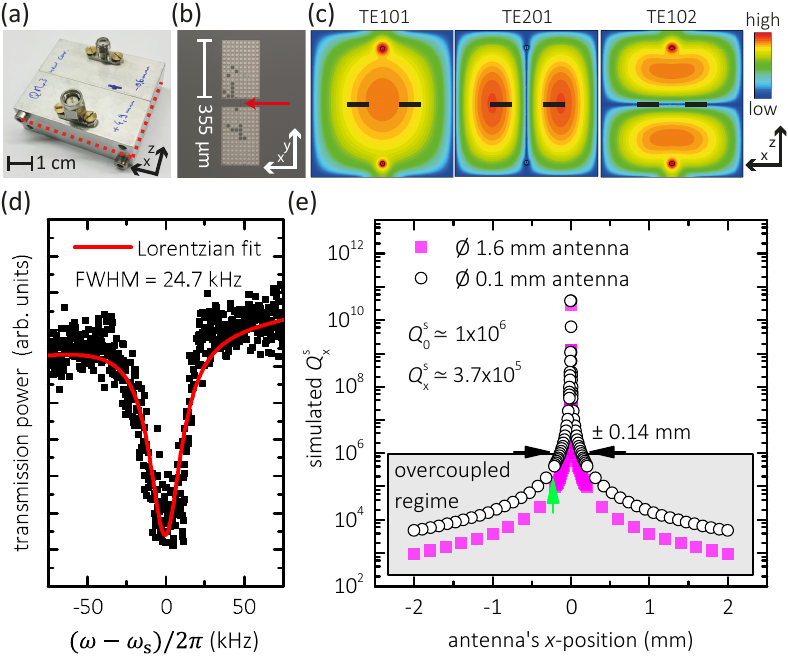}
\caption{(a) Photograph of the aluminum 3D cavity. The red dashed line indicates the equatorial plane of the cavity. (b) Optical micrograph of the superconducting transmon qubit. The two gridded metallic capacitors are connected via a submicron Josephson junction indicated by the red arrow. (c) Simulated $y$-component of the electrical field distribution in the cavity equatorial plane for the lowest three modes. Antenna pins are depicted as circles. The chip positions are marked with black rectangles. Their center positions are displaced in $x$-direction from the cavity center by \SI{6}{\milli\meter}. (d) Transmission power versus frequency of the storage mode (dots) with Lorentzian fit (line). (e) Simulated external $Q$-factor $Q_\mathrm{x}^\mathrm{s}$ of the storage mode for two different antenna diameters as a function of the antenna's $x$-position [$x=0$ denotes the cavity center as shown in (c)]. The overcoupled regime is marked with a grey box. The green arrow indicates $Q_\mathrm{x}^\mathrm{s}$ for this experiment (antenna diameter of \SI{0.1}{\milli\meter}). The black arrows show the positioning window to reach $Q_\text{x}^\mathrm{s} \geq Q_\text{0}^\mathrm{s}$. \label{fig:modes}}
\end{figure}

In this Letter, we introduce and analyze a compact quantum memory system based on a fixed-frequency transmon qubit\cite{Koch2007} coupled to two distinct modes of a single 3D microwave cavity (cf.\,Fig.\,\ref{fig:modes}). The cavity is made of aluminum (Al $\SI{99.5}{\percent}$) with two antennas for external coupling. It is mounted inside a dilution refrigerator with approximately \SI{30}{\milli\kelvin} base temperature. From a finite-element simulation \cite{CST} of the cavity transverse electric (TE) modes  [cf.\,Fig.\,\ref{fig:modes}\,(c)] we observe that the TE101 mode can be overcoupled while maintaining an undercoupled TE201 mode by properly placing the coupling antennas. In this way, the completely decoupled, highly coherent TE201 mode can be used for information storage and the well-coupled TE101 mode for dispersive readout at a megahertz rate \cite{Blais2004}. In our configuration, the fundamental TE101 readout mode $\omega_\text{RO}/2\pi = \SI{5.518}{\giga\hertz}$ is overcoupled and has a decay rate $\kappa_\text{RO}/2\pi = \SI{4}{\mega\hertz}$, whereas the first harmonic TE201 storage mode $\omega_\text{s}/2\pi = \SI{8.707546}{\giga\hertz}$ has a decay rate $\kappa_\text{s}/2\pi = \SI{24.7}{\kilo\hertz}$ [cf.\,Fig.\,\ref{fig:modes}\,(d)]. From a bare cavity measurement, we estimate the internal $Q$-factors $Q_\mathrm{0}^\mathrm{RO} \simeq \num{1.9e6}$ and $Q_\mathrm{0}^\mathrm{s} \simeq \num{1e6}$. For the TE201 mode to be symmetric with regard to the antennas, we insert two \SI{3x10}{\milli\meter} large silicon chips centered near the electric field antinodes. One serves as a dummy and on the other we fabricate the single-junction transmon qubit using aluminum technology and double-angle shadow evaporation. 
Due to its off-center placement inside the cavity, it couples to both the TE101 and the TE201 mode simultaneously with an equal coupling strength $g/2\pi\simeq\SI{53}{\mega\hertz}$. In addition, there is a small residual coupling $g_\text{102}/2\pi \simeq \SI{8}{\mega\hertz}$ to the TE102 mode\cite{Sears2012}. The qubit transition frequency $\omega_\text{q}/2\pi = \SI{6.234}{\giga\hertz}$ is designed to fall in between $\omega_\text{s}$ and $\omega_\text{RO}$ resulting in dispersive shifts of $\chi_\mathrm{RO} = \SI{3.6}{\mega\hertz}$ and $\chi_\mathrm{s} = \SI{1.1}{\mega\hertz}$ for the readout and storage mode, respectively. We measure an average qubit energy decay time $T_1^\mathrm{q} = \SI{1.32\pm0.05}{\micro\second}$ and a decoherence time $T_2^\mathrm{q} = \SI{2.49}{\micro\second}$, which is obtained via a Ramsey-type experiment. The qubit has an anharmonicity $\alpha = \omega_\mathrm{ef}-\omega_\mathrm{q} \simeq -2\pi\times\SI{185}{\mega\hertz}$. Due to the large detuning between qubit and readout mode, the qubit lifetime is not Purcell limited, even though the readout mode is designed to have a fast decay. For our sample, the Purcell limit for the qubit lifetime is calculated to be $\SI{20}{\micro\second}$.
\begin{figure}
\includegraphics[width=\columnwidth]{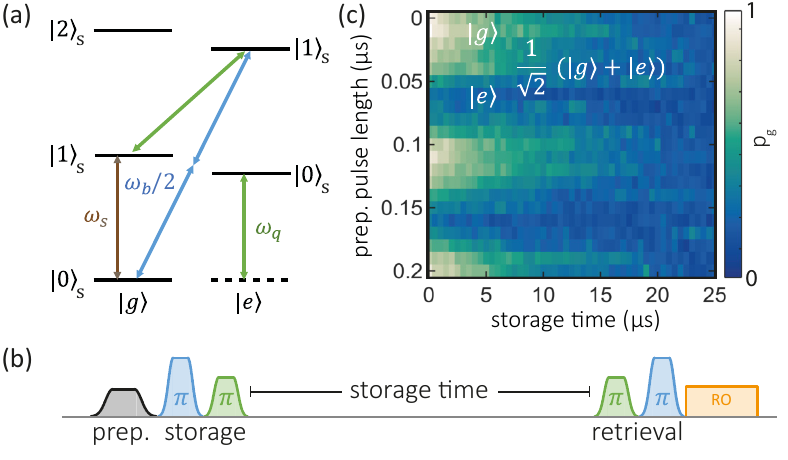}
\caption{(a) Energy level scheme of the qubit ($\omega_\text{q}$) coupled to the storage mode ($\omega_\text{s}$). We drive the second-order transition from $\left|g0\right\rangle$ to $\left|e1\right\rangle$ with two photons of frequency $\omega_\text{b}/2$. (b) Quantum memory protocol for storage and retrieval. Grey: Qubit preparation pulse. Blue: BSB pulse. Green: Qubit pulse. Orange: Readout (RO) pulse applied to the readout mode. (c) Decay of qubit superposition states transferred to the memory. The quantum memory protocol is applied to qubit states with different Rabi angles. We indicate three distinct prepared qubit states in the graph. An equal superposition state of the qubit is stored as $1/\sqrt{2}\left(\left|0\right\rangle_\mathrm{s}+\left|1\right\rangle_\mathrm{s}\right)$ in the memory.
\label{fig:protocol}}
\end{figure}

We now discuss the storage protocol based on the energy level diagram shown in Fig.\,\ref{fig:protocol}\,(a). Neglecting small second-order shifts, the equidistant ladder of the harmonic storage mode is shifted by the amount of the qubit energy if the qubit is excited. For the quantum memory protocol, we need to drive the blue sideband (BSB) transition from $\left|g0\right\rangle$ to $\left|e1\right\rangle$ between these two ladders with the frequency $\omega_\text{b} \equiv \omega_\text{s} + \omega_\text{q} + \chi_\text{s} + (2n_\mathrm{RO}-1) \chi_\text{RO}$. Here, we take into account the dispersive shifts $\chi_\mathrm{RO}$ of the readout and $\chi_\mathrm{s}$ of the storage mode. In the experiment, we keep $n_\mathrm{RO} \simeq 0$ during the protocol sequence. To fulfill parity conservation \cite{Goetz2017b}, we drive the BSB transition with two photons at $\omega_\text{b}/2$. This transition is described by the effective BSB Hamiltonian \cite{Blais2007,Wallraff2007}
\begin{equation}
\mathcal{H}_{\text{BSB}} = \frac{g^3 \Omega_\text{drv}^2}{(\omega_\text{s}-\omega_\text{b}/2)^2(\omega_\text{q}-\omega_\text{b}/2)^2} \left(\hat{a}^\dag\hat{\sigma}^+  + \hat{a} \hat{\sigma}^- \right)
\label{eq:bsb_ham}
\end{equation}
where $\Omega_\text{drv}$ is the drive amplitude, $\hat{a}^\dag (\hat{a})$ the creation (annihilation) operator of the storage mode, and $\hat{\sigma}^+ (\hat{\sigma}^-)$ the raising (lowering) operator of the transmon qubit. Our measurement protocol [cf.\,Fig.\,\ref{fig:protocol}\,(b)] starts with a qubit preparation pulse. By placing it at the beginning of the memory sequence, we are able to transfer various states into the memory.
The actual memory sequence begins with a $\pi$-pulse on the BSB transition. In this way, the ground state population of the qubit $p_\mathrm{g}$ is transferred to the first excited memory state on the right ladder. Then, by means of a qubit $\pi$-pulse, the qubit is deexcited and all population is swapped back to the left ladder. As a result, the qubit state is now encoded in the first two states of the storage mode \cite{Leek2010}. Specifically, the qubit ground (excited) state is transferred to the first excited (ground) state of the storage.
For state retrieval, we use the same pulse scheme in the reverse order and map the excitation back onto the qubit, which we read out dispersively. In general, we use flat-top Gaussian pulses with a rise time of $\SI{20}{\nano\second}$ to minimize leakage to higher qubit levels. For stability reasons, the AC-Stark shift of the qubit during the strong BSB drive is reduced by using filters (cf.\,Fig.\,S1 in Supplementary Material). In Fig.\,\ref{fig:protocol}\,(c), the decay during the storage time is shown for qubit states prepared with different preparation pulse lengths. The typical Rabi oscillation pattern is visible, however, with a significantly smaller decay rate as compared to that of the bare qubit. Hence, we find that superposition states of the qubit are stored as superpositions of Fock states in the storage mode as expected.
\begin{figure}
\includegraphics[width=\columnwidth]{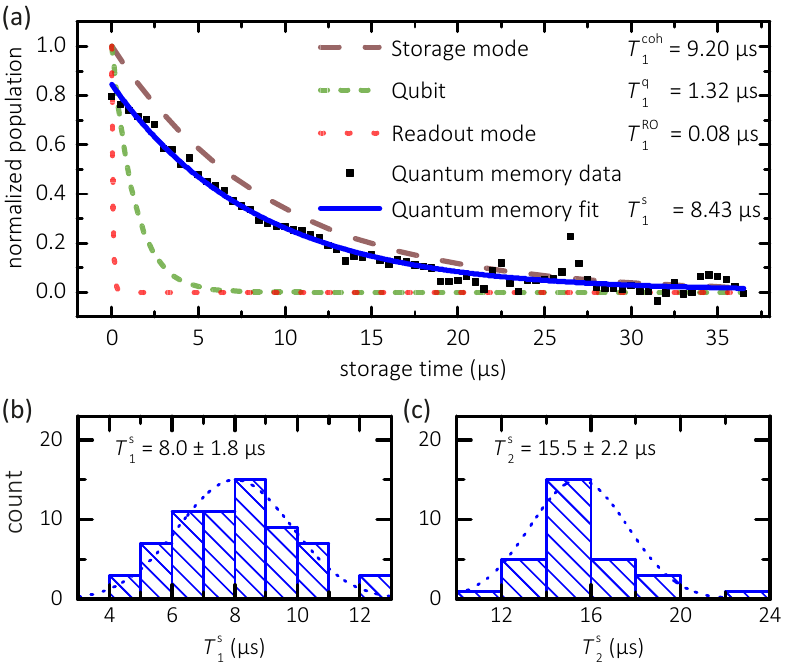}
\caption{(a) Comparison of decay rates of the quantum memory system. An exemplary trace of a single quantum memory measurement is shown (black rectangles). The corresponding decay time is obtained with an exponential fit. (b) and (c) Statistics of the $T_1^\mathrm{s}$ and $T_2^\mathrm{s}$ of a Fock state $\left|1\right\rangle_\mathrm{s}$ in the storage mode. 
\label{fig:T1_comparison}}
\end{figure}

In order to study the decay time of the memory, we use a protocol preparing the qubit in the $\left|g\right\rangle$ state to obtain the Fock state $\left|1\right\rangle_\mathrm{s}$ in the memory. The decay of this state is plotted versus the storage time in Fig.\,\ref{fig:T1_comparison}\,(a). We measure an average relaxation time $T_1^\mathrm{s} = \SI{8.0 \pm 1.8}{\micro\second}$. The data is in good agreement with an exponential decay. Small deviations from this behavior can be attributed to measurement artifacts. We can also perform a protocol preparing the qubit in state $1/\sqrt{2}\left(\left|g\right\rangle+\left|e\right\rangle\right)$ and rotating the qubit by $\pi/2$ at the end of the protocol. Such Ramsey-type measurement yields $T_2^\mathrm{s} = \SI{15.5\pm 2.2}{\micro\second}$. Figure\,\ref{fig:T1_comparison}\,(b) and (c) display histograms of the measured memory decay and decoherence times, respectively. Both follow a normal distribution. These fluctuations are frequently observed in literature and typically explained by ensembles of microscopic two-level fluctuators  \cite{Muller2015,Goetz2017}. The measured coherence time is slightly lower than expected from the $2T_1$-bound. As a result, we extract a dephasing time of $T_\varphi^\mathrm{s} = (1/T_2^\mathrm{s} - 1/2T_1^\mathrm{s})^{-1} = \SI{0.5}{\milli\second}$ for the storage mode. Originally, a bare harmonic oscillator exhibits no dephasing, however, in our case, we can explain this behavior by considering a stochastic qubit jump rate\cite{Rigetti2012} between $|g\rangle$ and $|e\rangle$. This results in frequency jumps of $\omega_\mathrm{s}$ due to the dispersive interaction and, as a consequence, a loss of phase information. We consider this mechanism to be dominant in our experiment because, via the relation \cite{Reagor2016} $\Gamma_\varphi \simeq P_\mathrm{e} \kappa_\mathrm{q}$, with $\kappa_\mathrm{q}$ being the qubit decay rate and $\Gamma_\varphi$ the pure dephasing rate, we extract a qubit excited state population in equilibrium $P_e = \SI{0.3}{\percent}$ in agreement with an estimate\cite{Jin2015} based on the temperature of the cavity walls of $\SI{50}{mK}$.
Furthermore, both $T_1^\mathrm{s}$ and $T_2^\mathrm{s}$ are by a factor of  6 longer than their bare-qubit counterparts. This observation is another indication, that the storage mode partially inherits the dephasing properties of the qubit. 
Finally, we compare the quantum storage time with the decay time of a coherent state in the storage mode. To this end, we excite the storage mode directly with a sufficiently long microwave pulse of frequency $\omega_\text{s}$ and directly monitor the field leaking out of the resonator (cf.\,Fig.\,S2 in Supplementary Material). The resulting value $T_1^\mathrm{coh} = \SI{9.2}{\micro\second}$ agrees well with the Fock-state storage time $T_1^\mathrm{s}$ within the expected statistical variation. 
The next parameter of interest is the decay time of the readout mode $T_1^\text{RO}$, which we determine to $T_1^\mathrm{RO} = \SI{80}{\nano\second}$ in a similar way.
This values shows that the overcoupled antenna configuration allows us to read out the qubit state on a time scale 100 times shorter than the storage time. A similar 2D system\cite{Leek2010} with a single planar on-chip resonator reaches only a factor of $37$. With a single 3D cavity\cite{Gasparinetti2016,Flurin2017}, ratios of $16.7$ and $154$ have been demonstrated. Using a combination of two rectangular superconducting 3D cavities \cite{Sirois2015,Vlastakis2013,Liu2017,Wang2017,Ofek2016}, where one serves for readout and the other one for storage, ratios of $15.8$, $45.8$, $1500$, $1818$ and $2500$ are obtained. By proper surface treatment of the cavity, we expect to be able to increase $Q_0^\mathrm{s}$ to approximately $\SI{5e7}{}$ as demonstrated in Ref.~\onlinecite{Reagor2013}. Our simulations~[cf.\,Fig.\,\ref{fig:modes}(e)] show that comparable values can be obtained for $Q_\mathrm{x}^\mathrm{s}$ with a more accurate antenna positioning of $\pm\SI{0.016}{\milli\meter}$. In this way, we predict that the ratio between readout and storage time could be extended to $\SI{25000}{}$, exceeding the highest currently reported value of 7300 for superconducting qubit-memory architectures employing cylindrical 3D cavities\cite{Rosenblum2017}. Other memory systems, such as spin-ensembles coupled to superconducting qubits, exhibit storage times comparable to rectangular superconducting 3D cavities\cite{Kubo2011,Grezes2015}. However, the spread in Rabi frequencies is a source for dephasing and the low cooperativity leads to a low efficiency in terms of the absorbed and re-emitted signal. Both drawbacks are not present in our type of system. Quantum memories based on nano-mechanical resonators also have similar storage times\cite{Reed2017}, but currently suffer from access times which are two orders of magnitude higher than those presented in this work. 

\begin{figure}
\includegraphics[width=\columnwidth]{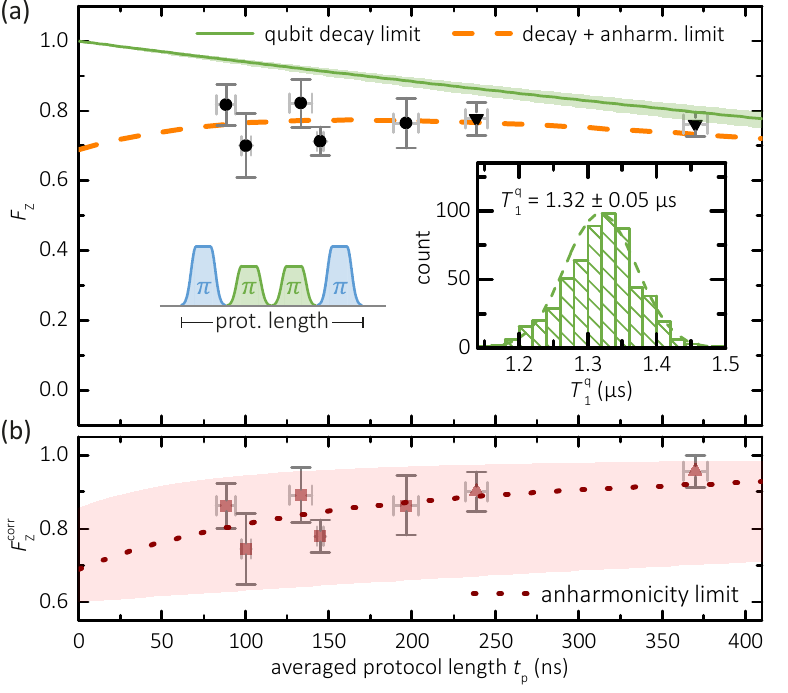}
\caption{(a) $Z$ fidelity measurements of the quantum memory protocol versus protocol length. Green line: qubit decay limit with $T_1^\text{q} = \SI{1.32 \pm 0.05}{\micro\second}$. Light green area: Statistical variations. Protocol lengths (inset) are set using different drive powers. Triangles: Long protocol lengths are achieved using $3\pi$ pulses for the qubit transition. Orange dashed line: product of the qubit decay limited fidelity and the anharmonicity limited fidelity $F_\mathrm{Z}^\mathrm{corr}$. Inset: Histogram of $T_1^\text{q}$. (b) $Z$ fidelity corrected for the qubit decay in order to make different protocol lengths comparable. Dotted line: fit. Red shaded region: fit uncertainty. 
\label{fig:zfid}}
\end{figure}

To complete the analysis of our quantum memory experiment, we characterize the fidelity of the protocol by means of the retrieved qubit population $p_\mathrm{g}(t_p)$ after the protocol of duration $t_p$. The $Z$ fidelity $F_\text{Z} = p_\mathrm{g}(t_\mathrm{p})/p_\mathrm{g}(0)$, which is the fidelity obtained from measurements along the quantization axis, is depicted in Fig.\,\ref{fig:zfid}\,(a) for seven different working points of the quantum memory protocol. At each working point, we perform several quantum memory measurements with calibrated $\pi$-pulses. The working points differ in the average protocol length $t_\mathrm{p}$. When using short pulses, which require a higher drive power to accomplish a $\pi$-rotation, we achieve a maximum of $F_\text{Z} = \SI{82\pm7}{\percent}$. Longer $t_\mathrm{p}$ let the maximum fidelity drop to $\SI{76\pm3}{\percent}$ due to qubit decay during the pulse sequence. However, within the statistical error margins, $F_\mathrm{Z}$ is constant over the range of protocol lengths. As a cross-check, we perform quantum process tomography of the memory protocol starting in the $\left|g0\right\rangle$ state. We find a process fidelity $F_\mathrm{QPT} = \SI{78}{\percent}$ (for technical details cf.~Supplementary Material). This value coincides very well with the outcomes of the $Z$ fidelity measurements and proves that our experiment is not limited by dephasing. Hence, further analysis is performed using $F_\mathrm{Z}$.
To shed more light onto the origin of these observations, we make $F_\mathrm{Z}$  from protocols of various lengths comparable by computationally eliminating the qubit decay. From this corrected $Z$ fidelity $F_\text{Z}^\text{corr} = F_\text{Z}/\exp(-t/T_1^\mathrm{q})$, we determine a maximum value of $F_\text{Z}^\text{corr} = \SI{96\pm4}{\percent}$ for $t_\mathrm{p} = \SI{370}{\nano\second}$ as shown in Fig.\,\ref{fig:zfid}\,(b). For shorter procotol lengths, we observe a lower $F_\text{Z}^\text{corr}$. This result points to a typical source of error for transmon qubits, namely state leakage caused by the low anharmonicity. To quantify the leakage \cite{Chen2016}, we use a simplified picture and define the leakage rate as $\Gamma_\text{L} = \gamma^\uparrow + \gamma^\downarrow + \gamma_\mathrm{sp}$, with the excitation rate $\gamma^\uparrow$, the stimulated emission rate $\gamma^\downarrow$ and the spontaneous emission rate $\gamma_\mathrm{sp}$ to and from higher states, respectively. Then, we can write the steady state leakage population $P_\text{L} = \gamma^\uparrow/\Gamma_\mathrm{L} \;[1-\exp(-\Gamma_\text{L} t_\text{p})] = a/(2a+\gamma_\mathrm{sp} t_\mathrm{p})\;[1-\exp(-2a-\gamma_\mathrm{sp} t_\text{p})]$, with $a = \gamma^\uparrow t_\mathrm{p} = \gamma^\downarrow t_\mathrm{p}$. The  rates $\gamma^\uparrow,\gamma^\downarrow$ are directly proportional to the drive strength $\Omega_\text{drv}$ and indirectly proportional to $t_\text{p}$, because the spectral width of our flat-top Gaussian shaped pulses with a fixed rise time is independent of the plateau length. We can identify $a$ with the ability of the drive pulse to create a leakage population, which depends on the fixed pulse rise time and the anharmonicity of the qubit. In the end, the corrected $Z$ fidelity is expected to behave as ${F}_\text{Z}^\text{corr} = 1-P_\text{L}(t_\mathrm{p}, a,\gamma_\mathrm{sp})$.
We use this function with $a$ and $\gamma_\mathrm{sp}$ as fit parameters to fit our data. Fitting the data we obtain the reasonable value $\gamma_\mathrm{sp}/2\pi = \SI{13.8 \pm 11.5}{\mega\hertz}$ for the spontaneous emission rate. Therefore, when aiming for even shorter pulses, the fidelity will be limited to $F_\text{Z}\simeq \SI{69}{\percent}$ due to the qubit anharmonicity. 
To further increase the fidelity, optimal control pulses are necessary to reduce state leakage for short pulses\cite{Motzoi2009}. In addition, a longer qubit lifetime \cite{Rigetti2012} would allow for longer pulses and, in turn, for a higher fidelity at a given anharmonicity.  

In conclusion, we have demonstrated a quantum memory protocol for a fixed-frequency transmon qubit by harnessing the multimode structure of a single 3D cavity. In this way, we have access to a long-lived storage mode but retain a fast readout capability. We are able to store qubit superposition states into the memory. The ratio between the readout and the storage rate shows a significant improvement compared to planar multimode resonators and is on a par with other measured values for single rectangular 3D cavities\cite{Flurin2017}. Furthermore, our measurements indicate that the dephasing of the memory is limited by the coupling to the qubit. We find a maximum $Z$ fidelity of $F_\mathrm{Z}= \SI{82\pm7}{\percent}$ limited by qubit relaxation and anharmonicity. This value matches well with the quantum process fidelity of $F_\mathrm{QPT} = \SI{78}{\percent}$. With few straightforward technical improvements, our compact platform can be further improved to store cat states\cite{Vlastakis2013} or GHZ states \cite{Paik2016} for continuous variable quantum computing. Regarding scalability, there are several attractive options. For example, more superconducting qubits using the storage mode as a bus can be inserted into the cavity for building logical qubits \cite{Gambetta2017}. Furthermore, coupling the qubit to higher cavity modes allows for a quantum memory register \cite{Kyaw2015,Naik2017}. With respect to analog quantum simulation, the implementation of Bose-Hubbard chains \cite{Leib2010} with the storage modes acting as bosons is facilitated by our compact architecture and less constraints on the amount of control lines. In particular, one can think of 2D lattice geometries for simulation of molecules \cite{Aspuru-Guzik2005} or solid state systems \cite{Buluta}.


\section*{Supplementary Material}
See supplementary material for details on sample fabrication, experimental techniques and quantum process tomography. 

\begin{acknowledgments}
We acknowledge  financial support from the German Research Foundation through FE 1564/1-1, the excellence cluster “Nanosystems Initiative Munich” (NIM), the doctorate program ExQM of the Elite Network of Bavaria, and the International Max Planck Research School “Quantum Science and Technology”.
\end{acknowledgments}

\bibliography{qm_bibfile}

\end{document}